\documentclass[twocolumn,showpacs,preprintnumbers,amsmath,amssymb,superscriptaddress]{revtex4}

\usepackage{graphicx}
\usepackage{dcolumn}
\usepackage{bm}
\usepackage{helvet}
\usepackage{gensymb}
\usepackage{textcomp}
\usepackage{subfiles}
\usepackage{hyperref}
\usepackage{xcolor}
\usepackage[figurename=Figure]{caption}
\usepackage{caption}
\usepackage{siunitx}
\hypersetup{
    colorlinks=true,
    linkcolor={blue},
    filecolor={red},      
    urlcolor={blue},
    citecolor={red},
    pdftitle={Overleaf Example},
    pdfpagemode=FullScreen,
    }
    
\begin{document}

\title{An Improved Paralyzable Detector Model}

\author{Yueyun Chen}
\affiliation{Department of Physics and Astronomy, University of California, Los Angeles, Los Angeles, California 90095, USA}
\affiliation{California NanoSystems Institute (CNSI), University of California, Los Angeles, California 90095, USA}

\author{Matthew Mecklenburg}\email{mmecklenburg@cnsi.ucla.edu}
\affiliation{California NanoSystems Institute (CNSI), University of California, Los Angeles, California 90095, USA}

\date{\today}

\begin{abstract}
Certain radiation detectors are `paralyzed' with high input count rates. When applied to count rates close to the event discriminator working rate the one-parameter dead time model fails. Here we present a corrected paralyzable detector model accounting for the event discriminator's finite response time. This two-parameter analytical model, when compared to the experimental data from a commercial x-ray detector, gives an improved description of the input and output count rate relations. Furthermore, it can independently determine the discriminator response time and the pulse shaper dead time, critical parameters for understanding a detector's performance. Finally, this model also provides a post-acquisition pile-up correction that greatly reduces artifacts in high-throughput spectra. In some situations, applying this model to optimize the acquisition and post-acquisition correction allows a user to acquire data an order of magnitude faster without compromising accuracy.
\end{abstract}

\maketitle

The precision of radiation counting experiments improves with increasing count number. Increasing count rates is the most important route towards improving signal-to-noise ratios (SNR). However, high count rates can lead to artifacts due to pile-up \cite{newbury_performing_2015}. The optimal data acquisition parameters represent a compromise between maximizing data throughput while minimizing artifacts \cite{hodoroaba_method_2014,newbury_performing_2015}. A precise understanding of the detector's behavior under varying count rate is essential for achieving this optimization.

\begin{figure}[h]
	\begin{center}
		\includegraphics[width=0.45\textwidth]{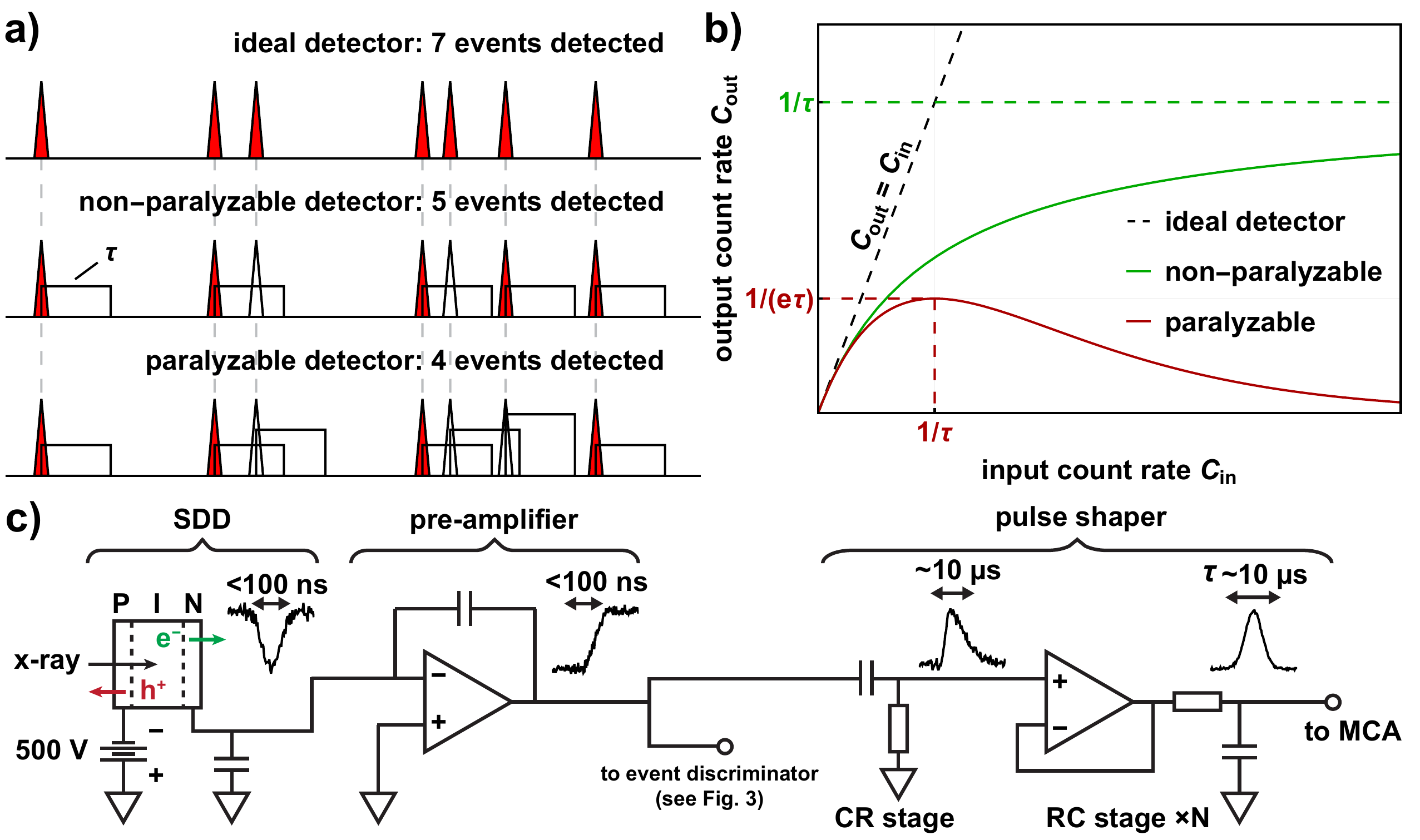}
		\caption{\textbf{Fundamental models illustration and radiation detection system schematic.} a) The same event sequence arrives at different types of detectors. The red and white triangles represent detected events and missed coincidence events, respectively. The rectangular box represents the recovery time (the dead time $\tau$). b) Output count rate $C_{\text{out}}$ as a function of input count rate $C_{\text{in}}$ for different types of detectors. The green dashed line marks the $C_{\text{out}}$ asymptote of the non-paralyazble detector, and the red dashed lines mark the position of maximal $C_{\text{out}}$ of the paralyzable detector. c) A schematic of a typical EDS system, where the progression of signal and approximate time constant is illustrated as an inset after each stage \cite{goldstein_x-ray_2003,ballabriga_design_2009}. This schematic can be generalized to other radiation detection systems by substituting the SDD with other types of detectors, such as neutron and electron detectors.}
		\label{fig:IdealModels}
	\end{center}
\end{figure}

With an ideal detector, one that measures every incident radiation quantum (Fig.~\ref{fig:IdealModels}a, top), a user can increase the input flux without limit to boost the data throughput. However, real-world detectors take a certain amount of time to measure a detected event. Sequential events occurring during the measurement period might not be counted, or might interfere with the first measurement. This measurement time is called the detector dead time $\tau$ \cite{leo_techniques_1994}. Two fundamental dead time models, the paralyzable \cite{evans_atomic_1955} and the non-paralyzable \cite{feller_probability_1948} detector models, handle the dead time differently. In the non-paralyzable detector model, events arriving during the dead time are simply not counted (Fig.~\ref{fig:IdealModels}a, middle). As a result, the output count rate $C_\text{out}$ increases monotonically as the input count rate $C_\text{in}$ increases. In the paralyzable detector model, on the other hand, events arriving during the dead time are not counted  and also reset the dead time (Fig.~\ref{fig:IdealModels}a, bottom), leading to a decrease in $C_\text{out}$ when $C_\text{in}$ passes a certain limit (Fig.~\ref{fig:IdealModels}b). Thus, high input fluxes can ``paralyze'' the detector.

Radiation events arrive at a detector in a discrete and random manner. For a non-paralyzable detector, the average measurement time per output event is the sum of the average arrival interval ($1/C_{\text{in}}$) and the detector dead time ($\tau$). So, the output count rate $C_{\text{out}}$ can be written as 
\begin{equation}
    C_{\text{out}} = 1/(1/C_{\text{in}} + \tau) = C_{\text{in}}/(1 + \tau C_{\text{in}} )
    \text{.} \label{eq:ideal_nonparalyzable}
\end{equation}
As $C_{\text{in}}$ increases, $C_{\text{out}}$ eventually approaches an asymptote where $(C_{\text{out}})_{\text{max}} = 1/\tau$ (Fig.~\ref{fig:IdealModels}b).

For a paralyzable detector, one event will be recorded when no other events arrive within a dead time $\tau$, and that probability is given by $P=e^{-\tau C_{\text{in}}}$ following Poisson statistics. So, the relationship between input and output count rate is 
\begin{equation}
    C_{\text{out}} = C_{\text{in}} e^{-\tau C_{\text{in}}}
    \text{.} \label{eq:ideal_paralyzable}
\end{equation}
Fig.~\ref{fig:IdealModels}b shows that $C_{\text{out}}$ of a paralyzable detector saturates very quickly as $C_{\text{in}}$ increases. The maximum occurs at $\tau C_{\text{in}} = 1$, where $(C_{\text{out}})_{\text{max}} = C_{\text{in}}/e$. Consider the dead time percentage $D_\%$ defined in Eq.~\ref{eq:deadtimepct},
\begin{equation}
    D_\% = (1 - C_{\text{out}}/C_{\text{in}}) \times 100\%
    \text{,} \label{eq:deadtimepct}
\end{equation}
it is easy to find that $C_{\text{out}}$ of a paralyzable detector is maximized at $D_\% = 1 - 1/e \approx 63.2$\% \cite{chen_measuring_2025}. This simple and dead-time independent relationship is used as a rule to optimize EDS acquisition throughput.

\begin{figure}[b]
	\begin{center}
		\includegraphics[width=0.45\textwidth]{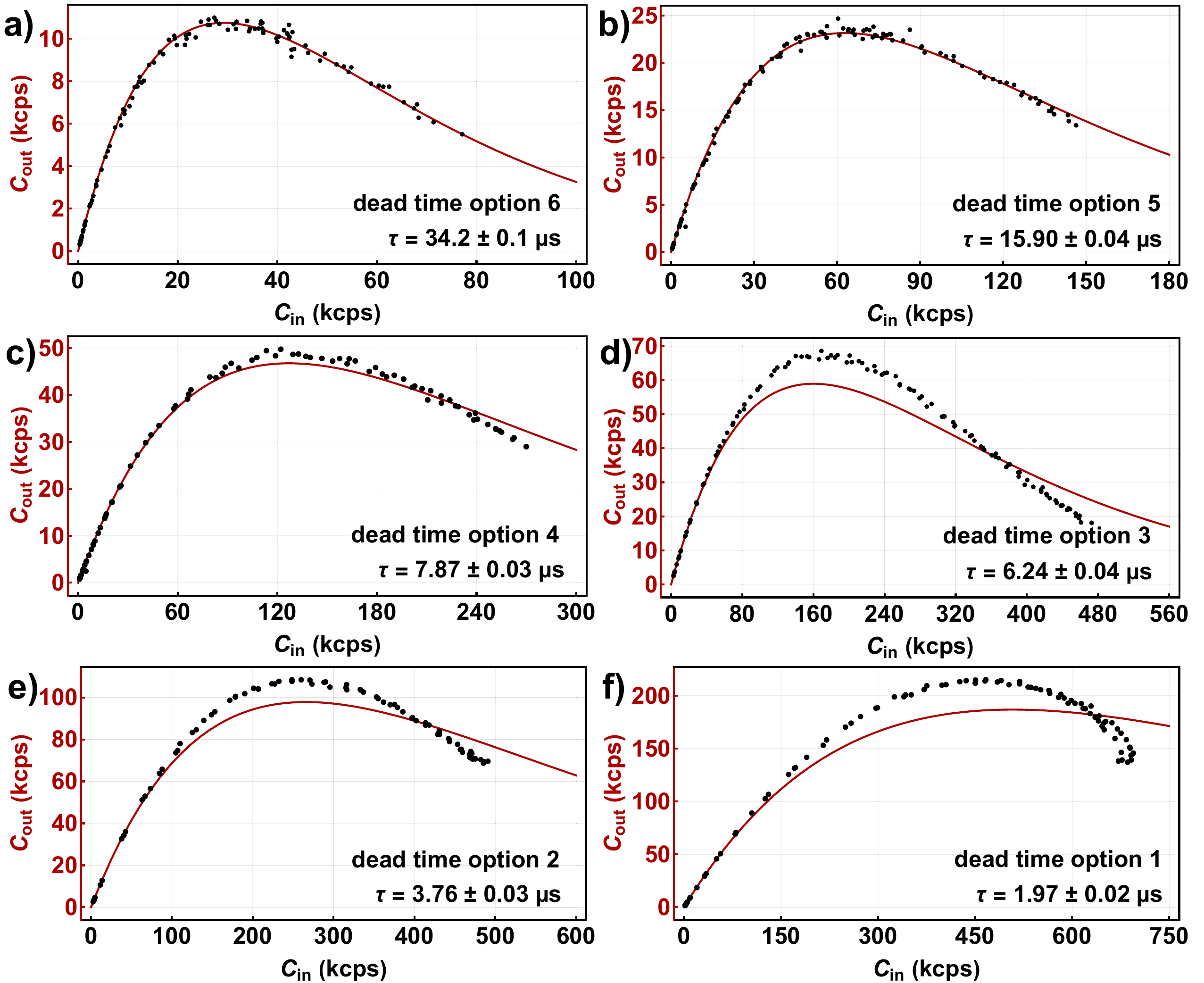}
		\caption{\textbf{Paralyzable detector model fits.} The experimental data (black dots) and corresponding fits (red curves) using the paralyable detector model (Eq.~\ref{eq:ideal_paralyzable}). The data is obtained on an Oxford X-MaxN 100TLE EDS detector for its six dead time options, and the measured dead time is labeled at the bottom right corner of each plot. The classic paralyzable detector model cannot accurately represent the data of shorter dead times.}
		\label{fig:ParalyzableModelFit}
	\end{center}
\end{figure}

Energy dispersive x-ray spectroscopy (EDS) is commonly used for elemental composition analysis in electron microscopes \cite{newbury_performing_2015}. Recently, it has been shown that with enough counts, EDS can have high enough precision to detect sub-eV chemical shifts \cite{chen_measuring_2025}. To achieve precise and accurate EDS measurements, spectra with high signal-to-noise ratios (SNR) and minimal artifacts are needed \cite{hodoroaba_method_2014,newbury_performing_2015}. However, the total count number needed might take hours to reach, which makes optimizing the data throughput while maintaining high data quality increasingly important. Having an analytic model that describes the behavior of the EDS detection systems precisely is beneficial. 

Contemporary EDS systems use silicon drift detectors (SDD) to measure x-ray energy \cite{lechner_silicon_1996}. The incoming photon is completely absorbed by the SDD and converted to a current pulse. The current is integrated by a capacitor (or a JFET) and converted to a voltage step through a charge sensitive pre-amplifier. The following pulse shaper minimizes the noise and converts the voltage step to a voltage pulse (Fig.~\ref{fig:IdealModels}c), whose height is proportional to the energy of the incident photon. The voltage pulse is then processed by a multi-channel analyzer (MCA) and added to the corresponding bin in the spectrum. However, the pulse shaper has a finite time constant, and only well-separated x-ray events can produce well-resolved pulses. Too high of an input rate can create too many pile-up pulses (multiple pulses merging into one). Pile-up pulses are rejected to avoid artifacts in the spectrum, and rejecting a massive amount of pile-up pulses can lead to a decrease in the number of output events. Thus, EDS detectors are paralyzable, with dead time $\tau$ dominated by the time constant of the pulse shaper \cite{goldstein_x-ray_2003}. A semi-Gaussian pulse shaper, which is a classic analog shaper with a short dead time and a high SNR \cite{noulis_integrated_2016,lutz_solid_2020}, is shown as an example in Fig.~\ref{fig:IdealModels}c. Other types of pulse shaper or digital signal processor can be implemented instead, but they are all subject to a finite dead time that limits the output rate \cite{nakhostin_introduction_2017}.

By fitting the $C_{\text{out}}$ versus $C_{\text{in}}$ data of an EDS system, the detector dead time can be experimentally determined. We recorded the input and output count rate of an Oxford X-MaxN 100TLE EDS detector on a FEI Titan 80-300 TEM. The count rates are taken from the Oxford AZtec v4.3 software. This detector features adjustable dead time $\tau$ (or process time, in the manufacturer's language) with six options. However, the exact values of $\tau$ are not publicly disclosed. In Fig.~\ref{fig:ParalyzableModelFit}, the data was fit with the paralyzable model (Eq.~\ref{eq:ideal_paralyzable}) to determine the value of $\tau$ for each dead time option. The paralyzable detector model works reasonably well for long dead times, but fails quickly as the dead time gets shorter. The failure of the model affects the rule of thumb 63.2\% dead time to maximize throughput. More importantly, it also suggests that the input count rate measured by the detector $C_{\text{hit}}$ deviates significantly from the actual input rate $C_{\text{in}}$ at shorter dead times. This deviation means that the pulse processor is not able to distinguish all input x-ray events, leading to a failure in hardware pile-up rejection \cite{nakhostin_digital_2010}.

\begin{figure}[h]
	\begin{center}
		\includegraphics[width=0.45\textwidth]{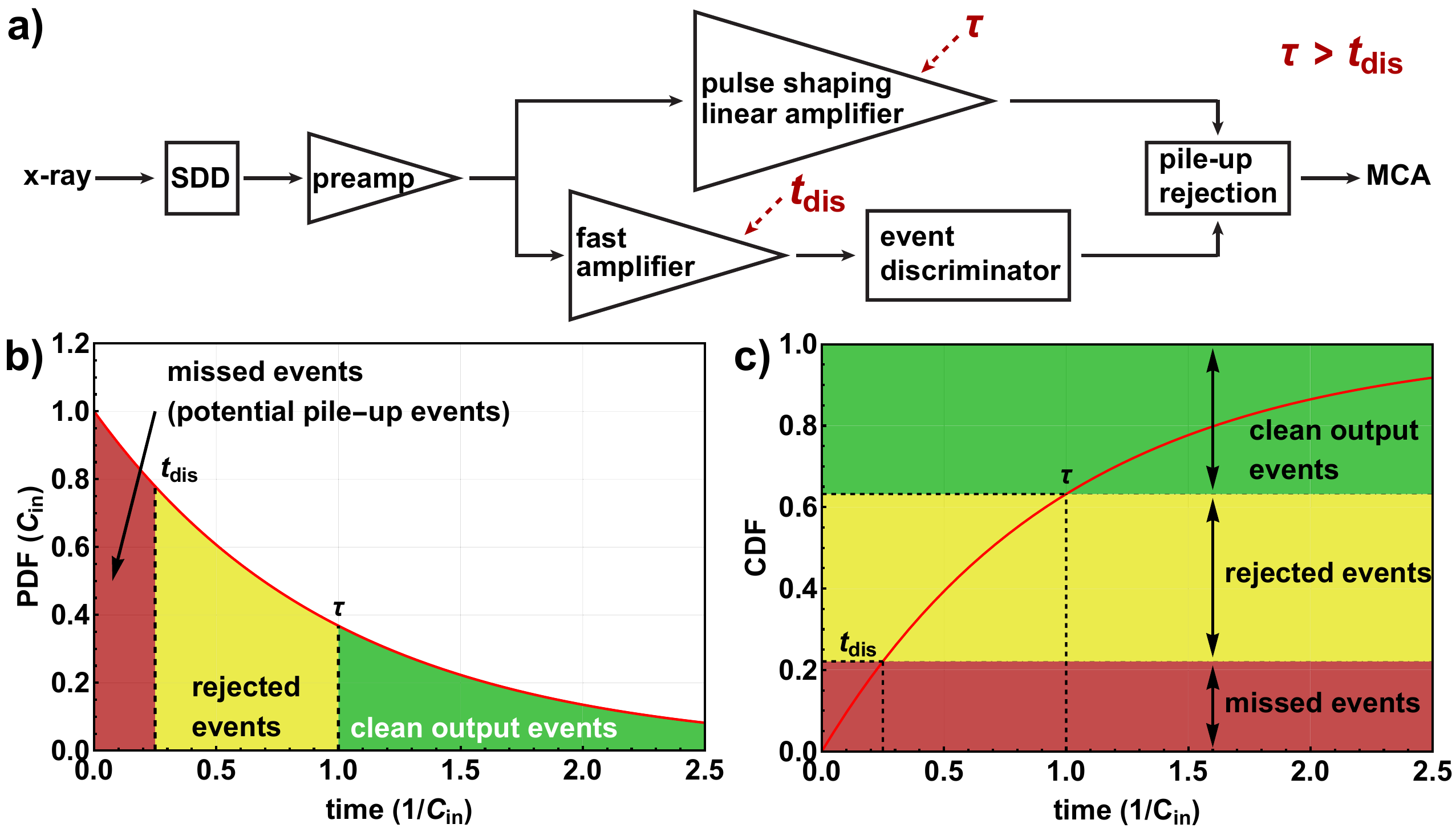}
		\caption{\textbf{EDS system with event discriminator.} a) Simplified schematic of an EDS system with event discriminator and hardware pile-up rejection \cite{goldstein_x-ray_2003}. The output of the pre-amplifier (pre-amp) goes into two parallel pathways, one with a pulse shaping linear amplifier used for energy measurement, and another one with a fast amplifier used for input event discrimination. The fast amplifier has a shorter response time $t_{\text{dis}}$, which is better for discriminating adjacent events in the expense of pulse height-energy linearity. b,c) Probability density function (PDF) and cumulative distribution function (CDF) of having a sequential event arriving within a time $t$. The time axis is in the unit of $1/C_{\text{in}}$. Sequential events arriving within the response time of event discriminator $t_{\text{dis}}$ are missed by the system and can become pile-up events that the system fails to reject. Sequential events arriving between $t_{\text{dis}}$ and the pulse shaping linear amplifier dead time $\tau$ are successfully rejected. If no sequential event arrives within $\tau$, the leading event becomes a clean output signal with no pile-up. For a complete event category diagram, refer to Fig.~\ref{fig:EventDiagram} in the appendix.}
		\label{fig:ProbabilityPlot}
	\end{center}
\end{figure}

To account for the inaccurately detected input count rate, it is helpful to introduce a finite response time $t_{\text{dis}}$ to model the input event discriminator of the system. The $C_{\text{out}}$ described in the following text relates to the more comprehensive model, not the ideal one above.  The event discriminator is a subsystem parallel to the pulse shaper (Fig.~\ref{fig:ProbabilityPlot}a). It has a faster response and is used to discriminate input events for hardware pile-up rejection \cite{goldstein_x-ray_2003}. Modeling the event discriminator as a paralyzable detector following Eq.~\ref{eq:ideal_paralyzable}, the relationship between the measured input count rate $C_{\text{hit}}$ and the actual input count rate $C_{\text{in}}$ is given by
\begin{equation}
    C_{\text{hit}} = C_{\text{in}} e^{-t_{\text{dis}} C_{\text{in}}}
    \text{.} \label{eq:detected_input}
\end{equation}
To illustrate how the detector sorts the input events into different categories, we plot the probability density function (PDF) and the cumulative distribution function (CDF) of having a sequential event arriving within a time $t$ after a leading event (Fig.~\ref{fig:ProbabilityPlot}b,c). The area under the PDF can be divided into three regions: the green region indicates the probability that the previous event is well resolved and becomes a clean output event, 
\begin{equation}
    P(\text{clean})=e^{-\tau C_{\text{in}}}
    \text{,} \label{eq:prob_clean_output}
\end{equation}
the yellow region indicates the probability that the previous event is correctly rejected due to a subsequent pile-up event identified by the input event discriminator, and the red region indicates the probability that the following event is missed due to the finite response time of the input discriminator 
\begin{equation}
    P(\text{missed})=1-e^{-t_{\text{dis}} C_{\text{in}}}
    \text{.}\label{eq:prob_miss}
\end{equation}
When a subsequent event is missed and no other events are detected between $t_{\text{dis}}$ and $\tau$, the previous event becomes a pile-up output event. The probability of an event becoming an output event is 
\begin{equation}
    P(\text{out})=\frac{P(\text{clean})}{1- P(\text{missed})}=e^{-(\tau-t_{\text{dis}}) C_{\text{in}}}
    \text{.} \label{eq:prob_output}
\end{equation}
The probability of an event becoming a pile-up pulse missed by the hardware pile-up rejection system and ending up in the output spectrum is 
\begin{align}
    P(\text{pile-up}) &= \frac{P(\text{missed})}{1-P(\text{missed})} \, P(\text{out})\\
    &= e^{-(\tau-t_{\text{dis}}) C_{\text{in}}}(e^{t_{\text{dis}} C_{\text{in}}}-1)
    \text{.} \label{eq:prob_pile-up}
\end{align}
A more detailed description of how events are classified into different categories is given in Appendix A.

Assuming that we stay on the ascending branch of the input event discriminator (see Fig.~\ref{fig:IdealModels}b and Eq.~\ref{eq:ideal_paralyzable}, $t_{\text{dis}} C_{\text{in}} < 1$), then we can solve Eq.~\ref{eq:detected_input} for $C_{\text{in}}$,
\begin{equation}
    C_{\text{in}} = -\frac{1}{t_{\text{dis}}} W_0 (-t_{\text{dis}} C_{\text{hit}})
    \text{,}\label{eq:actual_input}
\end{equation}
where $W_0$ is the principal branch of the Lambert $W$ function, which has been shown to be useful for inverting the classic paralyzable detector model \cite{heemskerk_gamma_2020}. Combining Eq.~\ref{eq:ideal_paralyzable} and Eq.~\ref{eq:prob_output}, we can obtain the total output rate $C_{\text{out}}$ as a function of the measured input rate $C_{\text{hit}}$
\begin{equation}
    C_{\text{out}}= -\frac{1}{t_{\text{dis}}} W_0 (-t_{\text{dis}} C_{\text{hit}}) e^{\frac{1}{t_{\text{dis}}} W_0 (-t_{\text{dis}} C_{\text{hit}} ) (\tau-t_{\text{dis}})}
    \text{.} \label{eq:corrected_model}
\end{equation}
Eq.~\ref{eq:corrected_model} provides a model corrected for a system with a real-world event discriminator and hardware pile-up rejection, which is the most common configuration for contemporary EDS systems \cite{goldstein_x-ray_2003}. This total output can be separated from the clean (pulse pile-up free) output by $(\tau-t_{\text{dis}})\rightarrow\tau$. Eq.~\ref{eq:corrected_model} can be further simplified to 
\begin{equation}
    C_{\text{out}} = C_{\text{hit}} e^{\frac{1}{t_{\text{dis}}} W_0 (-t_{\text{dis}} C_{\text{hit}} ) (\tau - 2t_{\text{dis}})}
    \text{,} \label{eq:corrected_model_simplified}
\end{equation}
using the Lambert identity $W(z)e^{W(z)}=z$.

\begin{figure}[h]
	\begin{center}
		\includegraphics[width=0.45\textwidth]{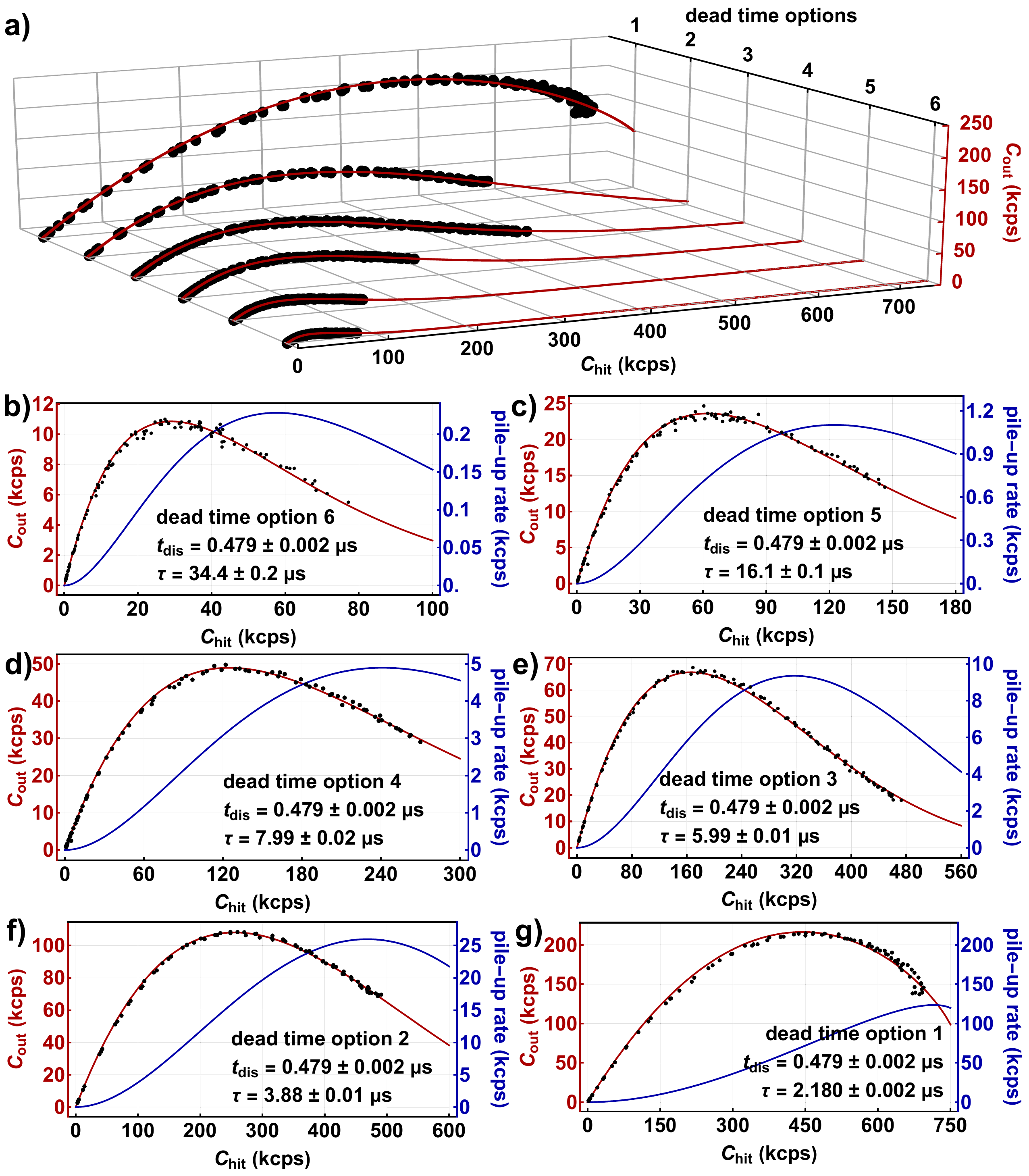}
		\caption{\textbf{Fits using model corrected for a real-world event discriminator.} Experimental data in Fig.~\ref{fig:ParalyzableModelFit} is fitted again using the corrected model in Eq.~\ref{eq:corrected_model_simplified} which accounts for an event discriminator and pile-up rejection system. a) A simultaneous fit to all six dead time options to determine the common parameter $t_{\text{dis}}$. It is assumed that all six dead time options use the same event discriminator. b-g) $C_{\text{out}}$ vs $C_{\text{hit}}$ fit (red) and predicted pile-up rate (blue) for each dead time option. Compared to the classic paralyzable detector model in Fig.~\ref{fig:ParalyzableModelFit}, the corrected model accurately represents the data even to the shortest dead time.}
		\label{fig:CorrectedModelFit}
	\end{center}
\end{figure}

We perform a simultaneous fit to determine the response time $t_{\text{dis}}$ of the event discriminator and six different dead times (Fig.~\ref{fig:CorrectedModelFit}a) using the corrected model. The fitted curve and data for each dead time option are plotted separately in Fig.~\ref{fig:CorrectedModelFit}b-g. Compared to Fig.~\ref{fig:ParalyzableModelFit}, even though $\tau$ only changes by about 5\%, the corrected model agrees much better with the data. The response time of the event discriminator is determined to be $0.479 \pm 0.002$~\micro s, which matches our expectation (between the $\sim$100~ns SDD charge collection time and the \micro s level pulse shaper dead time).

Being a generalized model, the corrected paralyzable detector model (Eq.~\ref{eq:corrected_model_simplified}) provides a way to measure the response time of the event discriminator and the dead time of EDS system without knowing the detailed structure of each system component. It also provides a new way to measure the dead time of radiation detectors with similar structures, without the need for two standard radiation sources (two-source method) \cite{almutairi_simultaneous_2021}. More importantly, it helps to maximize throughput more accurately and offers the ability to quantify the pulse pile-up effect, which negatively impacts the spectrum quality. Pulse pile-up generates false peaks that can be misidentified as elements do not exist in the specimen, for example the sum peak. It can also lead to shifts in peak energy and changes in peak intensity, which introduce errors in elemental composition quantification \cite{newbury_performing_2015,statham_pile-up_2006}. Thus, it is helpful to quantify the rate of pile-up events for evaluating spectrum quality. The rate of pile-up events that enter the output spectrum is 
\begin{align}
    C_{\text{pile-up}} &= C_{\text{in}} \times P(\text{pile-up}) \\ 
     &= C_{\text{in}} e^{-(\tau-t_{\text{dis}}) C_{\text{in}}}(e^{t_{\text{dis}} C_{\text{in}}}-1)
    \text{,} \label{eq:pileup_rate}
\end{align}
where $C_{\text{in}}$ is the actual input count rate, but can be written as a function of the measured input rate $C_{\text{hit}}$ with Eq.~\ref{eq:actual_input}. The blue curves in Fig.~\ref{fig:CorrectedModelFit}b-g are the pile-up event count rate predicted by Eq.~\ref{eq:pileup_rate}. Note that $C_{\text{pile-up}}$ can go higher than $C_{\text{out}}$ in principle, because multiple pile-up events can go into the same output event. But $C_{\text{pile-up}} > C_{\text{out}}$ only happens when you go far beyond the optimal data acquisition condition.

\begin{figure}
	\begin{center}
		\includegraphics[width=0.40\textwidth]{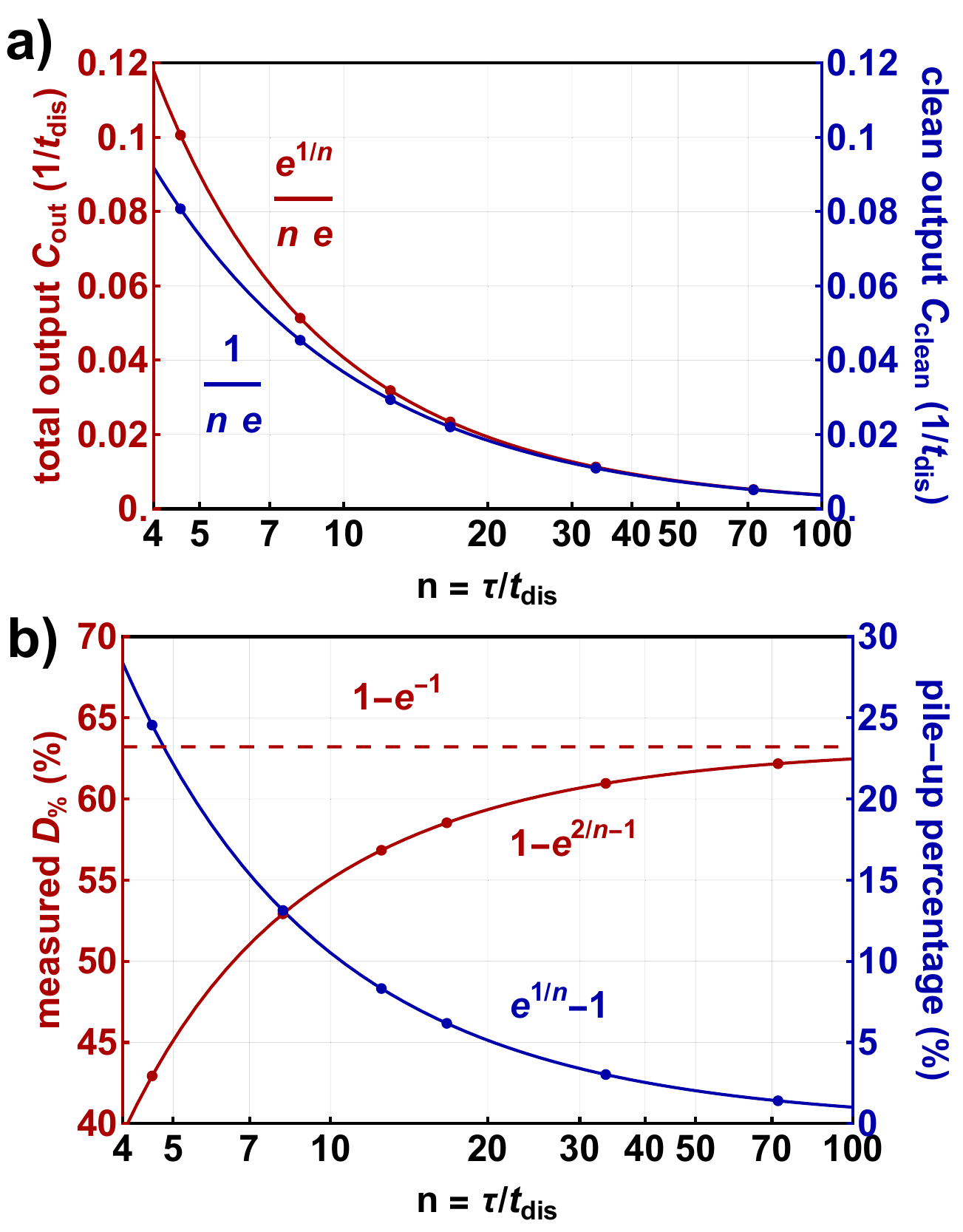}
		\caption{\textbf{Maximizing clean events throughput.} Maximizing the count rate of clean output events as a function of the ratio of dead time $\tau$ and the response time of the event discriminator $t_{\text{dis}}$. The dots along the curves represent the ratios measured on our Oxford X-MaxN 100TLE EDS detector for different dead time options. These plots guide users to maximize the data throughput and estimate the pile-up effect in the acquired spectrum. The $D_\%$ corresponding to the max output rate deviates more from the prediction of the classic paralyzable detector model (dashed line) as the ratio $n$ gets smaller. Note that the measured $D_\%$ given in Eq.~\ref{eq:measured_deadtimepct}, and the dimensionless formulas can be derived with the probabilities shown in Appendix A.}
		\label{fig:PileupPlot}
	\end{center}
\end{figure}

The optimal data acquisition condition is when the clean output rate $C_\text{clean} = C_\text{in} \times P(\text{clean})$ is maximized (Eq.~\ref{eq:prob_clean_output}). Fig.~\ref{fig:PileupPlot} shows a few related parameters as a function of the ratio of dead time $\tau$ and the response time of the event discriminator $t_{\text{dis}}$ under the optimal conditions. The measured dead time percentage,
\begin{equation}
    \text{measured}\ D_\%  = (1 - C_{\text{out}}/C_{\text{hit}}) \times 100\%
    \text{,}\label{eq:measured_deadtimepct}
\end{equation}
is plotted here instead of the actual $D_\%$ in Eq.~\ref{eq:deadtimepct}, because commercial software is usually unaware of the difference between $C_{\text{hit}}$ and $C_{\text{in}}$ and reports the measured $D_\%$ to users. The smaller the ratio $n$, the more the measured $D_\%$ under the optimal condition (red curve in Fig.~\ref{fig:PileupPlot}b) deviates from the ideal ($t_{\text{dis}}\rightarrow 0$) 63.2\% (dashed line) and the higher pile-up percentage (number of pile-up events divided by the total output events, or $P(\text{pile-up})/P(\text{out})\times 100\%$). However, being able to quantify how many pile-up events are added to the spectrum provides a method to apply post-acquisition correction, which can largely eliminate pile-up artifacts and maintain the accuracy of EDS measurements even with ten times higher throughput. Post-acquisition pile-up correction methods have been developed by detector manufacturers \cite{statham_pile-up_2006,eggert_automated_2011,eggert_smart_2012}, but the algorithm details are often not made public, and the corrected spectra are often not accessible for further analysis outside of the commercial software. 

\begin{figure}
	\begin{center}
		\includegraphics[width=0.40\textwidth]{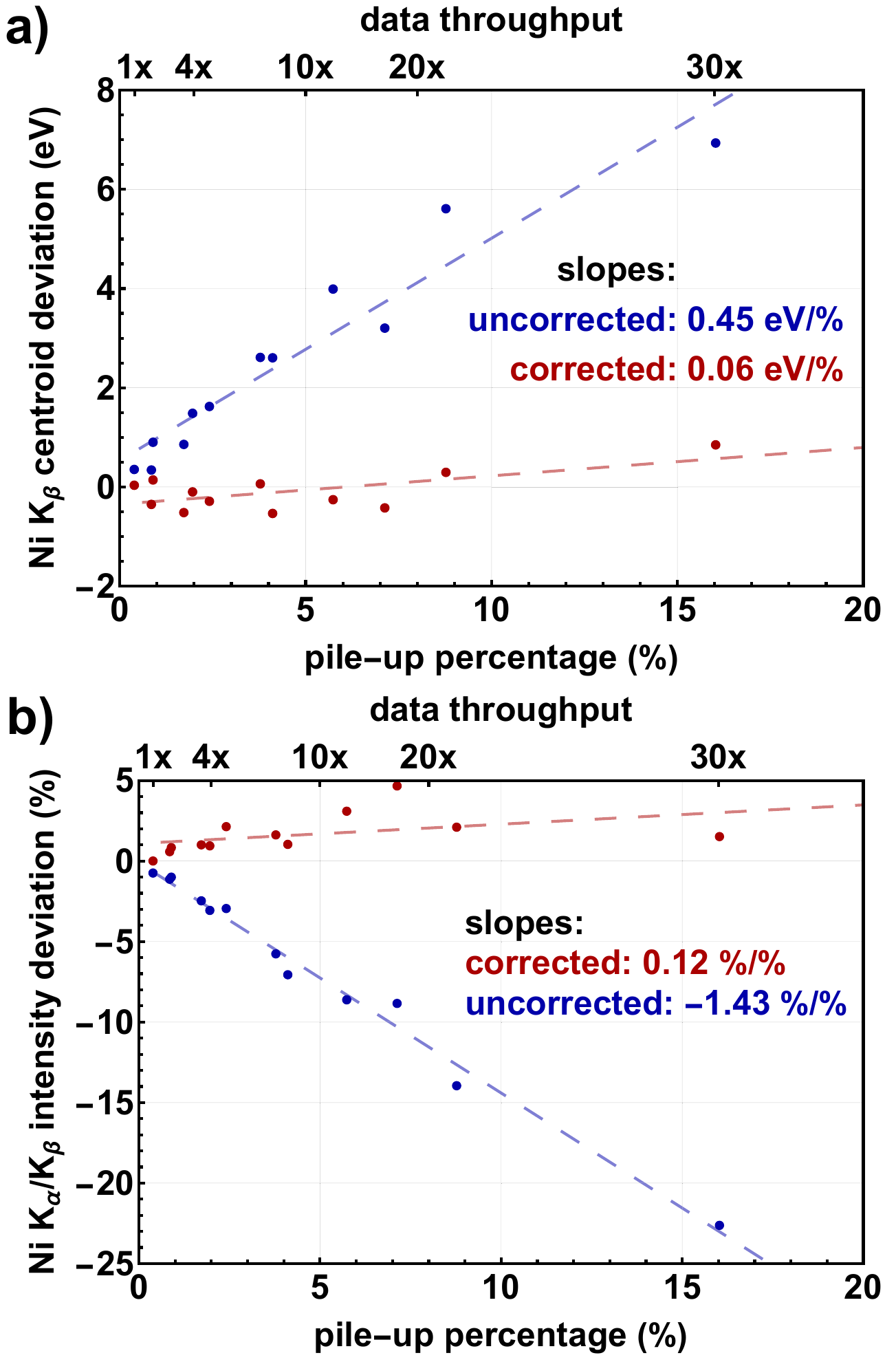}
		\caption{\textbf{Pile-up correction allows accurate quantification with high data throughput.} Pile-up correction significantly reduces the deviation of EDS peak energy a) and peak intensity b) as a function of pile-up percentage or data throughput. The workflow presented, determining the detector dead time using the corrected paralyzable detector model and applying pile-up correction, provides a way to acquire data with high throughput while maintaining the quantification accuracy. The data throughput is not directly proportional to the pile-up percentage (it also depends on the dead time used), the scale in the plot is determined by a few selected spectra.}
		\label{fig:PileupCorrection_Plot}
	\end{center}
\end{figure}

Here we demonstrate a post-acquisition pile-up correction algorithm based on the corrected paralyzable detector model (with details in appendix B) and apply it to both simulated (Fig.~\ref{fig:PileupCorrection_Simulation}) and experimental (Fig.~\ref{fig:PileupCorrection_Experiment}) NiO$_\text{x}$ EDS spectra. Knowing the measured input rate $C_{\text{hit}}$ and the output rate $C_{\text{out}}$, this algorithm can remove two-photon coincidence events that the hardware failed to reject. To quantify the improvement from pile-up correction, we focus on the Ni~K$_\beta$ peak that is skewed by three pile-up peaks nearby (Fig.~\ref{fig:PileupCorrection_Simulation} and \ref{fig:PileupCorrection_Experiment}). By measuring the peak energy of Ni~K$_\beta$ (Fig.~\ref{fig:PileupCorrection_Plot}a) and the intensity ratio of Ni~K$_\alpha$/K$_\beta$ (Fig.~\ref{fig:PileupCorrection_Plot}b) of 12 experimental spectra acquired under different input x-ray fluxes and different dead time options, we show that the post acquisition pile-up correction can provide an order of magnitude improvement in quantification accuracy for the spectra acquired with high data throughput.

In summary, we present a corrected paralyzable detector model for particle detectors that accounts for an event discriminator and hardware pile-up rejection system. Applying this model to an Oxford X-MaxN 100TLE EDS detector, we reach excellent agreement with the input and output count rate data. The new model allows users to determine the response time $t_{\text{dis}}$ of the event discriminator and the dead time $\tau$ of the pulse shaper, which are often not published by the detector manufacturer, but are critical for optimizing data throughput. More importantly, the corrected paralyzable detector model provides a way to implement post-acquisition pile-up correction, which is essential to achieve high data throughput while maintaining high quantification accuracy. Knowing the output count rate $C_{\text{out}}$ (or the measured input count rate $C_{\text{hit}}$), the event discriminator response time $t_{\text{dis}}$, and the dead time $\tau$, two-photon pile-up peaks can be effectively removed. This pile-up correction largely reduces peak misidentification and significantly improves the accuracy of atomic percentage and peak energy quantification. 

The data analysis pipeline, from quantifying detector dead time to post-acquisition pile-up correction, demonstrates that the data throughput can be increased by more than ten times while maintaining a sub-eV level peak energy accuracy needed for EDS chemical shift measurement. With a state-of-the-art EDS detector whose output rate can exceed a million counts per second, the data acquisition time for a chemical shift measurement can be reduced from hours to minutes, making it more practical to perform on a daily basis \cite{chen_measuring_2025}. Although using a shorter dead time to increase data throughput often leads to broadening of EDS peaks (due to strobe peak broadening), this is less of an issue in the high energy range where EDS excels (due to the domination of the Fano factor term, see Eq.~\ref{eq:fwhm}). Moreover, this corrected paralyzable detector model is not unique to x-ray silicon drift detectors, but generally applicable to particle detectors that measure energy spectrum with single particle sensitivity (eg.\ HPGe gamma-ray spectrometer \cite{hafizoglu_efficiency_2024} and scintillator-based SiPM detector \cite{buonanno_gamma-ray_2023}).

This work was supported by the BioPACIFIC Materials Innovation Platform of the National Science Foundation under Awards No. DMR-1933487 and DMR-2445868, and NSF STC award DMR-1548924 (STROBE). Data were collected at  the Electron Imaging Center for Nanosystems (EICN) (RRID:SCR$\_$022900) at the University of California, Los Angeles’s California for NanoSystems Institute (CNSI). The authors would like to thank Dr. B. C. Regan from Department of Physics and Astronomy, UCLA, for his insightful comments on the manuscript.

\appendix 
\counterwithin{figure}{section}
\section{Event Classification}
\begin{figure}[h]
	\begin{center}
		\includegraphics[width=0.45\textwidth]{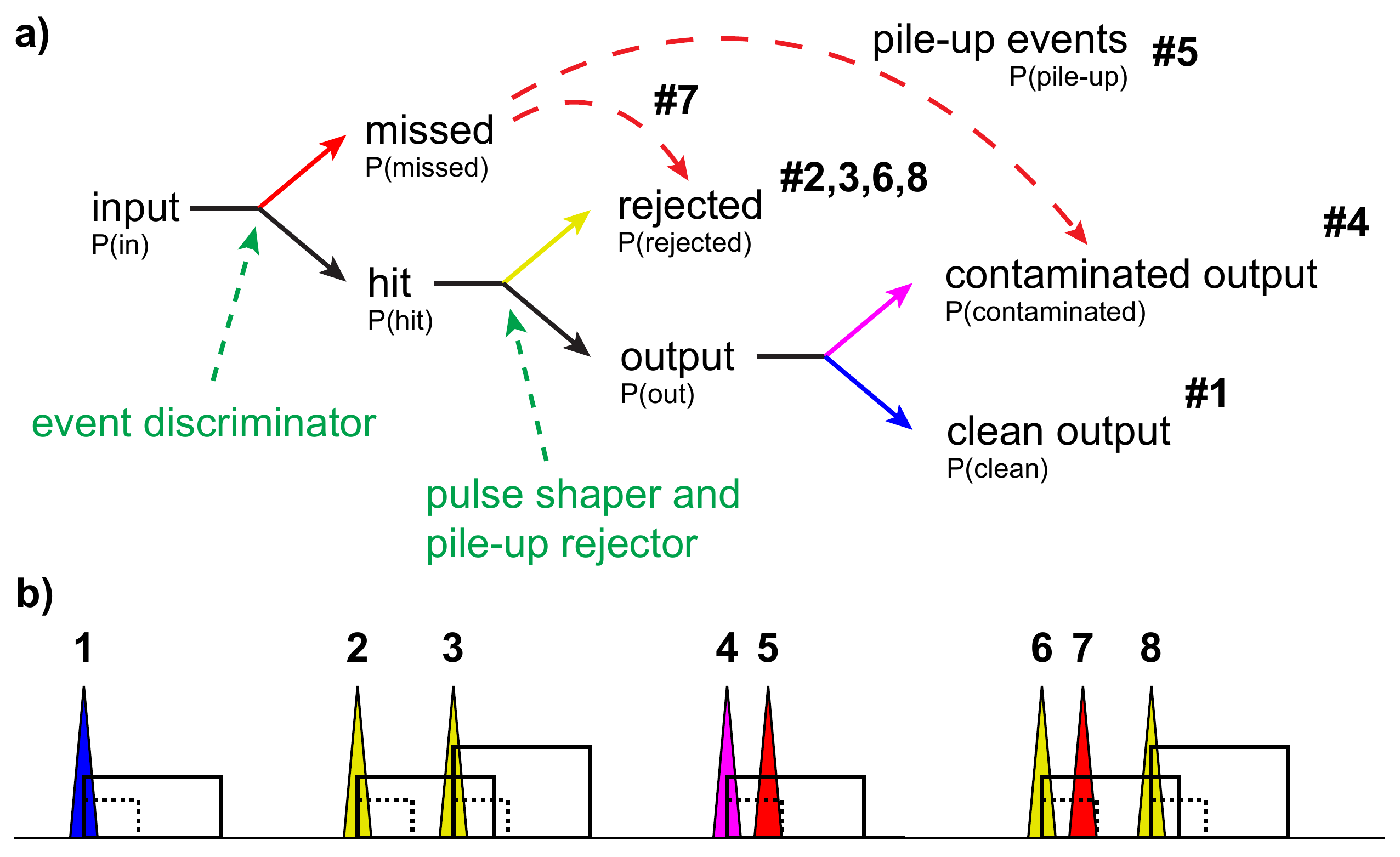}
		\caption{\textbf{Event diagram.} a) The x-rays reaching the detector can be classified basing on how they are treated by the event discriminator and the pile-up rejection system. The events passed both layers and reached the output spectrum can be further separated as clean outputs and outputs contaminated by pile-up events. An event missed by the event discriminator always overlaps with a hit (detected) event closely. If this hit event passes the pile-up rejector and ends up in the output spectrum, then it becomes a contaminated output, and the overlapping missed event becomes a pile-up event. The total energy of events 4 and 5, summed, will appear in the detector's output. b) An event sequence similar to the ones in Fig.~\ref{fig:IdealModels}a. They are color coded to indicate which category they end up being in the diagram above, and their indices are also labeled next to where they belong. The solid and dashed line boxes indicate the pulse shaper dead time $\tau$ and the event discriminator response time $t_{\text{dis}}$, respectively.}
		\label{fig:EventDiagram}
	\end{center}
\end{figure}

Understanding how x-ray events are classified by the hardware is essential for quantifying pile-up artifacts and developing a post-acquisition pile-up correction algorithm. The times of arrival of input events are first registered by the event discriminator. During this process, events arriving within time $t_{\text{dis}}$ following a previous event are missed by the event discriminator (Fig.~\ref{fig:EventDiagram}b \#5, 7). They sneak into the pulse processing system without being registered and can potentially become pile-up events. If a second event is detected within time $\tau$ following a registered event, then the pile-up rejector will discard both to prevent overlap (Fig.~\ref{fig:EventDiagram}b \#2, 3, 6, 8). If no event is detected within time $\tau$, then the first event will be sent to the output spectrum (Fig.~\ref{fig:EventDiagram}b \#1, 4). In the case where a missed event follows the output event, the output event is contaminated by pile-up (Fig.~\ref{fig:EventDiagram}b \#4). 

The probability of an input event ending up in each category can be derived as a function of $\tau$, $t_{\text{dis}}$, and $C_{\text{in}}$:

\begin{equation}
    P(\text{in}) = 1
\end{equation}

\begin{equation}
    P(\text{hit}) = e^{-t_{\text{dis}} C_{\text{in}}}
\end{equation}

\begin{equation}
    P(\text{missed}) = 1 - P(\text{hit}) = 1 - e^{-t_{\text{dis}} C_{\text{in}}}
\end{equation}

\begin{equation}
    P(\text{out}) = e^{-(\tau - t_{\text{dis}}) C_{\text{in}}}
\end{equation}

\begin{align}
    P(\text{rejected}) &= P(\text{hit}) - P(\text{out})\\
    &= e^{-t_{\text{dis}} C_{\text{in}}} - e^{-(\tau - t_{\text{dis}}) C_{\text{in}}}
\end{align}

\begin{equation}
    P(\text{clean}) = e^{-\tau C_{\text{in}}}
\end{equation}

\begin{align}
    P(\text{contaminated}) &= P(\text{out}) - P(\text{clean})\\
    &= e^{-\tau C_{\text{in}}}(e^{t_{\text{dis}} C_{\text{in}}}-1)
\end{align}

\begin{align}
    P(\text{pile-up}) &= P(\text{contaminated})/P(\text{hit}) \\
    &= e^{-(\tau - t_{\text{dis}}) C_{\text{in}}}(e^{t_{\text{dis}} C_{\text{in}}}-1)
\end{align}

\section{Pile-up Correction Algorithm}
Here we present a simple pile-up correction algorithm that corrects for two-photon coincidence events. It uses the characteristic times determined with the corrected paralyzable detector model and can be performed by the users themselves. This algorithm is based on probability and does not require any knowledge of the specimen. Knowing $t_{\text{dis}}$, $\tau$, and $C_{\text{out}}$, this post-acquisition correction can be performed numerically on any EDS spectrum with the help of the corrected paralyzable detector model described in this work. Consider an energy spectrum starting at 0~eV with $N$ bins. Pile-up effect always pushes photon events to a bin with higher energy. So, bin $k$ in the spectrum is only affected by photons with energy between 0~eV and the energy of bin $k-1$. Denote the uncorrected and corrected counts from bin $1$ to bin $k-1$ as $uc_{1:k-1}$ and $c_{1:k-1}$, respectively. Assuming a spectrum has been corrected from bin $1$ to bin $k-1$, then we can calculate the pile-up counts added to bin $k$ and subtract it off. The number of pile-up events that enter the spectrum with the corresponding energy is 
\begin{equation}
    pu_{1:k-1} = \left[ c_{1:k-1}/P(\text{clean}) \right]\times P(\text{pile-up})
    \text{.} \label{eq:puc_pileup_counts}
\end{equation}
Because the input is random, these pile-up events are randomly distributed to all output events. And to end up in bin $k$, a pile-up event with an energy corresponding to bin $m$ has to add to an event with an energy corresponding to bin $k-m$. So, the probability that these pile-up events contribute to bin $k$ is given by
\begin{equation}
    ap_{1:k-1} = \left[c_{k-1:1}/P(\text{clean})\right]\times P(\text{out})/\sum_{i=1}^{N} uc_i
    \text{.} \label{eq:puc_addin_prob}
\end{equation}
Since we are only correcting for two-photon coincidence events, we need to account for the probability that a pile-up event ends up as a two-photon coincidence event rather than a multi-photon coincidence event, which is 
\begin{equation}
    P(\text{two-photon}) = e^{-t_{\text{dis}} C_{\text{in}}}
    \text{.} \label{eq:puc_two_photon}
\end{equation}
Knowing Eq.~\ref{eq:puc_pileup_counts}, \ref{eq:puc_addin_prob}, and \ref{eq:puc_two_photon}, we can correct for the pile-up events in bin $k$ with Eq.~\ref{eq:puc_correction}.
\begin{equation}
    c_k = uc_k - (pu_{1:k-1} \cdot ap_{1:k-1}) \times P(\text{two-photon})
    \label{eq:puc_correction}
\end{equation}
Once bin $k$ is corrected, the same procedure can be applied iteratively to correct for bin $k+1$, and so on. Because the 0~eV bin is not affected by the pile-up effect ($c_1=uc_1$), it provides a starting point for this iterative algorithm. And iterating from $k=2$ to $k=N$ corrects the entire spectrum for two-photon coincidence events. 

\begin{figure}[h]
	\begin{center}
		\includegraphics[width=0.45\textwidth]{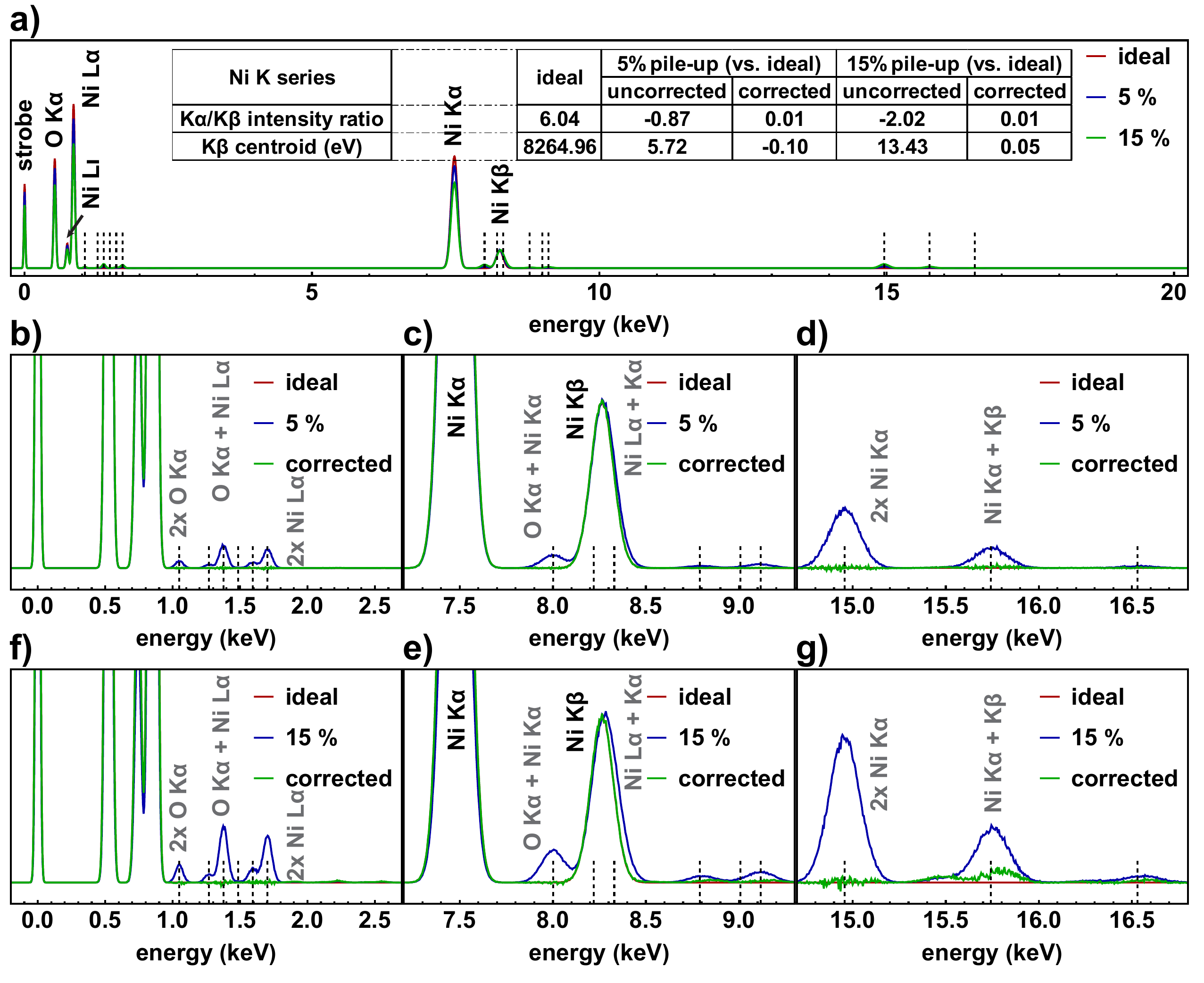}
		\caption{\textbf{Pile-up correction on simulated EDS spectra.} a) Simulated EDS spectra of NiO$_\text{x}$ with 0\% (ideal), 5\%, and 15\% pile-up events. The dashed lines indicate the location of all two-photon pile-up peaks. The Ni~K$_\beta$ peak at 8265~eV is affected by pile-up peaks. The inset table shows the deviations of Ni~K$_\alpha$/K$_\beta$ intensity ratio and Ni~K$_\beta$ centroid due to pile-up effects. The pile-up correction mostly removes the deviations are reduced by two orders of magnitude. b-d) Comparison between spectrum with 5\% pile-up before and after post-acquisition correction. All two-photon pile-up peaks are removed. f-g) Comparison between spectrum with 5\% pile-up before and after post-acquisition correction. All two-photon pile-up peaks are removed, with minor effects from higher order pile-up visible in \textit{g)}. All y-axes are counts. Several selected pile-up peaks are labeled with gray texts.}
		\label{fig:PileupCorrection_Simulation}
	\end{center}
\end{figure}

In the end, a normalization factor can be applied to effectively put all the coincidence events back into the correct bin (Eq.~\ref{eq:puc_normalization}). This normalization factor does not affect EDS quantification but is applied to all corrected spectra shown here for the ease of comparison. 
\begin{equation}
    n_0 = \left[P(\text{out}) + P(\text{pile-up})\right]/P(\text{clean})
    \label{eq:puc_normalization}
\end{equation}
In most of the equations above, the actual input rate $C_{\text{in}}$ is directly referenced for simplicity. Although $C_{\text{in}}$ is not measurable, it can be derived from $t_{\text{dis}}$, $\tau$, and $C_{\text{out}}$ using Eq.~\ref{eq:actual_input} and \ref{eq:corrected_model}. In the case where the spectrum contains negative energy bins to capture the entire strobe peak (electronic noise peak), the negative energy bins are ignored.

We first test the pile-up correction algorithm on simulated spectra (Fig.~\ref{fig:PileupCorrection_Simulation}). Consider a high-quality EDS spectrum of a NiO$_\text{x}$ standard, the most prominent peaks are the strobe peak, O~K$_\alpha$, Ni~L$_\iota$, Ni~L$_\alpha$, Ni~K$_\alpha$, and Ni~K$_\beta$ peaks. Five million events are distributed among these peaks with an intensity ratio of 2:4:1:6:12:2. The energies of the events are randomly generated following a normal distribution around each peak. The peak energy is obtained from the literature \cite{bearden_x-ray_1967,deslattes_x-ray_2003}, and the peak width at energy $\varepsilon_x$ is determined by Eq.~\ref{eq:fwhm} \cite{chen_measuring_2025}. The full width at half maximum (FWHM) of the strobe peak, the silicon electron-hole pair creation energy, and the silicon Fano factor are taken to be $\text{FWHM}_\text{strobe} = 30$~eV, $\varepsilon_i = 3.6$~eV \cite{lechner_pair_1996}, and $F = 0.12$ \cite{lowe_measurement_2007}, respectively. 
\begin{equation}
   \text{FWHM}(\varepsilon_x) = \sqrt{(2\sqrt{2\text{ln}2F\varepsilon_x\varepsilon_i})^2+\text{FWHM}_{\text{strobe}}^2}
    \label{eq:fwhm}
\end{equation}
A certain number of events are randomly selected to be pile-up events and randomly added to the rest of the events. The events are then sorted into a 4096-bin spectrum with 5~eV bin size and a start (end) energy of $-242.5$~eV ($+20242.5$~eV).

Three spectra with 0\% (ideal), 5\%, and 15\% pile-up are generated, and the locations of all two-photon pile-up peaks are indicated with dashed lines (Fig.~\ref{fig:PileupCorrection_Simulation}a). For an event discriminator response time of $t_{\text{dis}} = 0.5$~\micro s and a dead time of $\tau = 2$~\micro s (parameters similar to dead time option 1 on our Oxford X-MaxN 100TLE detector), 5\% and 15\% pile-up correspond to an actual input count rate $C_{\text{in}}$ of 98~kcps and 280~kcps, respectively. Pile-up corrections are then applied to the 5\% and the 15\% pile-up spectra. Comparisons between the ideal spectrum, the uncorrected spectrum, and the corrected spectrum in three different energy ranges are shown in Fig.~\ref{fig:PileupCorrection_Simulation}b-d for the 5\% pile-up case and in Fig.~\ref{fig:PileupCorrection_Simulation}e-g for the 15\% pile-up case. The pile-up peaks at 1.7~keV, 8~keV, and 15~keV can be misidentified as silicon K$_\alpha$ (1.74~keV  \cite{bearden_x-ray_1967}), copper K$_\alpha$ (8.05~keV  \cite{bearden_x-ray_1967}), and yttrium K$_{\alpha}$ (14.96~keV \cite{bearden_x-ray_1967}), respectively. Pile-up artifacts also change the Ni~K$_\alpha$/K$_\beta$ intensity ratio and shift the Ni~K$_\beta$ peak to the right. The pile-up correction algorithm effectively eliminates all two-photon pile-up peaks for both cases, and the corrected spectra overlap with the ideal spectrum within the noise level. Some three-photon pile-up peaks that are not taken into account can be seen in Fig.~\ref{fig:PileupCorrection_Simulation}g. The pile-up correction significantly reduces peak misidentification issue and would improve the accuracy of quantification analysis.

To quantify the improvement from pile-up correction, we focus on the Ni~K$_\beta$ peak, which is affected by three pile-up peaks nearby. Without pile-up correction, Ni~K$_\beta$ is shifted to the right. And if we calculate the centroid position, the shift is on the order of a few eVs (Fig.~\ref{fig:PileupCorrection_Simulation}a inset table). Such artifacts hugely impede the efforts to detect fine shifts in peak energy using EDS (chemical shift detection \cite{chen_measuring_2025}, for example), and force this kind of study to be performed with an ultra-low throughput to limit pile-up artifacts. However, after applying the pile-up correction, sub-eV level peak centroid accuracy is recovered even for the 15\% pile-up spectrum. Looking at the peak intensity ratio between the Ni~K$_\alpha$ peak and the Ni~K$_\beta$ peak, which should be constant, we also observe one to two orders of magnitude improvement in accuracy. The effectiveness of a simple post-acquisition pile-up correction algorithm shows that it is possible to ramp up the data throughput while maintaining a high accuracy for peak centroid (important for chemical shift study) and peak intensity (important for elemental composition study) determination. 

\begin{figure}
	\begin{center}
		\includegraphics[width=0.45\textwidth]{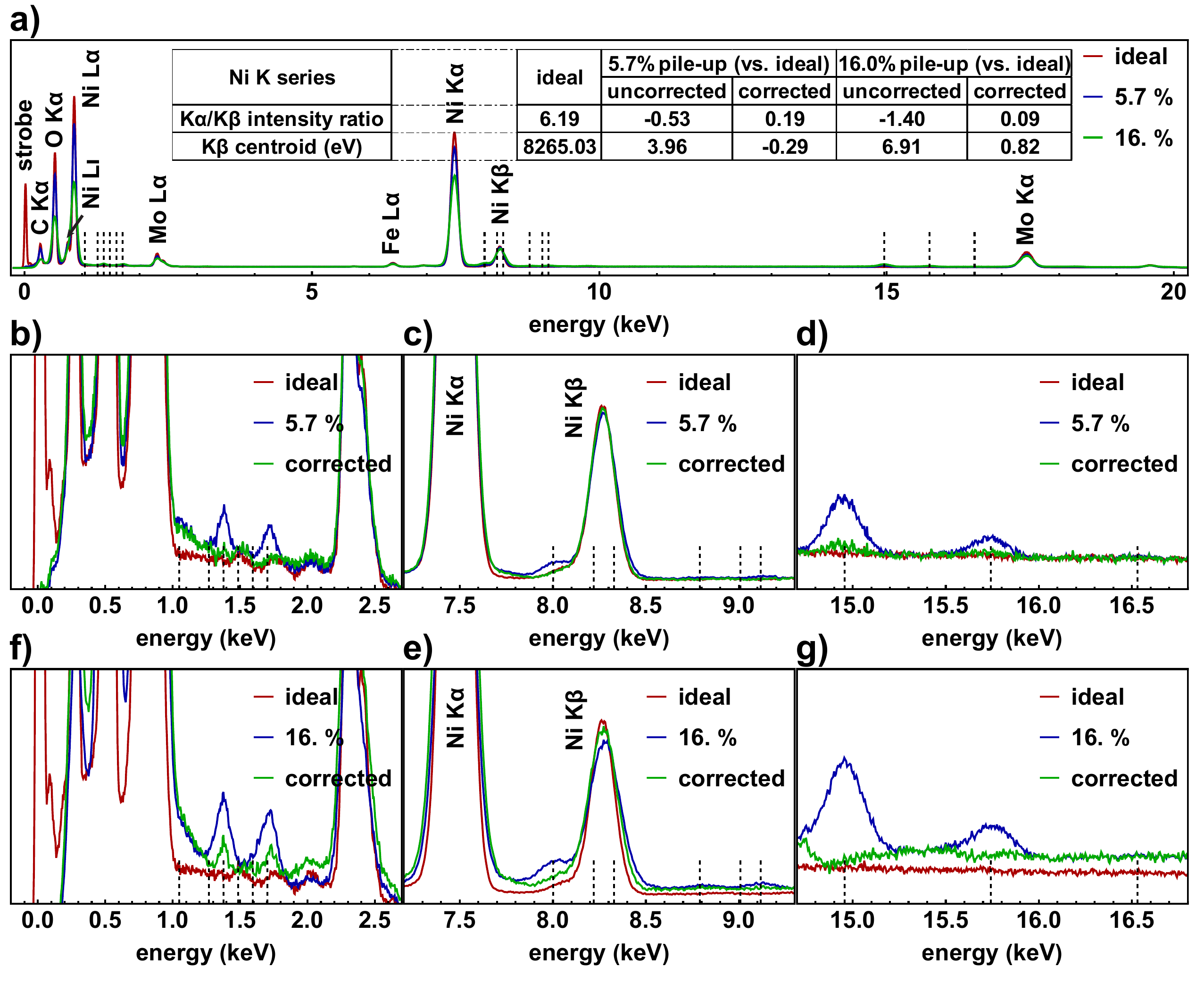}
		\caption{\textbf{Pile-up correction on experimental EDS spectra.} a) Experimental spectra acquired on a NiO$_\text{x}$ standard with similar pile-up percentage as simulated spectra shown in Fig.~\ref{fig:PileupCorrection_Simulation}. The extra EDS peaks of C, Mo, and Fe come from the carbon membrane, molybdenum TEM grid, and the TEM objective lens. The dashed lines indicate the location of the same two-photon pile-up peaks shown in Fig.~\ref{fig:PileupCorrection_Simulation}. b-d) Comparison of spectrum with 5.7\% pile-up before and after post-acquisition correction. f-g) Comparison of spectrum with 16\% pile-up before and after post-acquisition correction. The post-acquisition pile-up correction significantly reduces the intensity and centroid deviation of Ni~K$_\beta$ peak, and the peak energy can be determined with sub-eV accuracy even on spectrum with 16\% pile-up. All y-axes are counts. Refer to Fig.~\ref{fig:PileupCorrection_Simulation} for pile-up peak labels.}
		\label{fig:PileupCorrection_Experiment}
	\end{center}
\end{figure}

We also demonstrate that this pile-up correction routine works reasonably well on experimental spectra (Fig.~\ref{fig:PileupCorrection_Experiment}). To obtain spectra similar to the simulated ones, we use a NiO$_\text{x}$ standard (PELCO NiO$_\text{x}$ test specimen No.\ 650) and acquire twelve spectra with different dead time options and x-ray fluxes. The detector is the same Oxford X-MaxN 100TLE EDS detector, and each experimental spectrum contains five million counts. However, because of the continuous Bremsstrahlung X-ray background \cite{goldstein_x-ray_2003} taking about 40\% of the total counts, the experimental spectra are subject to more Poisson noise than the simulated ones. Pile-up percentage is calculated using $t_{\text{dis}}$ and $\tau$ determined in Fig.~\ref{fig:CorrectedModelFit}, and the pile-up correction is applied following Eq.~\ref{eq:puc_correction}. Three spectra with pile-up percentages similar to simulated ones are plotted in Fig.~\ref{fig:PileupCorrection_Experiment}. The ideal, 5.7\%, and 16\% pile-up spectra are acquired with dead time option 6 and $C_{\text{out}} = 5.83$~kcps, dead time option 3 and $C_{\text{out}} = 60.8$~kcps, and dead time option 1 and $C_{\text{out}} = 182$~kcps, respectively. The raw data of the ideal spectrum is acquired with a very low input x-ray flux to minimize pile-up events, and the pile-up percentage is estimated to be 0.4\%. The ideal spectrum shown in Fig.~\ref{fig:PileupCorrection_Experiment} already has the pile-up correction applied. The channel dispersion and offset are calibrated using the centroids of O~K$_\alpha$ and Ni~K$_\alpha$ peaks, the background intensity (determined with an energy window of 8.55--8.65~keV) is subtracted when quantifying Ni~K series peaks.

The post-acquisition pile-up correction effectively removes all visible pile-up peaks from the spectrum with 5.7\% pile-up (Fig.~\ref{fig:PileupCorrection_Experiment}b-d) and suppresses those in the spectrum with 16\% pile-up (Fig.~\ref{fig:PileupCorrection_Experiment}e-g). Despite some insufficient correction visible around 1.5~keV for the 16\% pile-up spectrum, the pile-up correction significantly reduces the intensity ratio deviation and the centroid deviation for both experimental spectra (Fig.~\ref{fig:PileupCorrection_Experiment}a inset). Sub-eV accuracy in Ni~K$_\beta$ peak centroid is preserved even for the spectrum with 16\% pile-up, whose data throughput is 30 times higher than the condition used for the acquisition of the ideal spectrum. The Ni~K$_\beta$ peak centroid and the Ni~K$_\alpha$/K$_\beta$ intensity ratio of all experimental spectra are shown in Fig.~\ref{fig:PileupCorrection_Plot}. Pile-up correction greatly improves the quantification accuracy for spectra acquired with high data throughput.

\bibliographystyle{apsrev4-1}
\bibliography{DetectorModel}

\end{document}